\documentstyle[11pt,aaspp4,psfig]{article}

\begin{document}

\hspace{3.5 in}  {\it Science}, {\bf 280}, 1405 (1998)
%\small
%\hspace{4in}CfPA preprint \#98-th-09
%\normalsize

\title{\large
\bf\sc Extracting Primordial Density Fluctuations}
%{\small
%\it Testing Models of Structure Formation in the Universe}}

\author{Eric Gawiser}

\affil {Department of Physics, University of California at Berkeley, 
Berkeley CA  94720}

\author{Joseph Silk}

\affil {Departments of Physics and Astronomy, and Center for Particle 
Astrophysics, University of California at Berkeley, Berkeley, CA  94720}
%{\it Fellow, Center for Particle Astrophysics}}
\authoremail{gawiser@astron.berkeley.edu}
%\author{\it Center for Particle Astrophysics}

%\begin{center}
%To appear in \textit{Science}
%\end{center}

\begin{abstract}

The combination of detections of anisotropy in the Cosmic Microwave Background 
radiation and observations of the large-scale distribution of galaxies  
probes the 
primordial density fluctuations of the universe on spatial 
scales varying by three orders of magnitude.
These data are found to be inconsistent with the predictions of 
several popular cosmological models.
Agreement between the data and the Cold + Hot 
Dark Matter model, however, suggests that a significant fraction of
the matter in the universe may consist of massive neutrinos.

\end{abstract}

\section{Introduction}

Shortly after the Big Bang,
the universe 
was
smooth to a precision of one part in $10^5$.  We measure this smoothness in 
the cosmic microwave background (CMB) radiation, 
the photons which provide us with a record of conditions in the 
early universe because they were 
last scattered about 300,000 years after the Big Bang.  
To remarkable precision, the early universe was
characterized by isotropic,
homogeneous expansion.  
However, temperature fluctuations
are measured in the CMB ({\it 1}), 
and complex structure surrounds us.  
There is a simple
connection;  the seeds of large-scale structure were infinitesimal density
perturbations that grew via gravitational instability into massive
structures such as galaxies and galaxy clusters.

One can search for the primordial seeds of large-scale structure by two
complementary techniques.  The cosmic microwave background fluctuations probe
 the density
fluctuations in the early universe on comoving scales greater than 
$\sim 100$ Mpc.
The gravity 
field of these density fluctuations also generates
fluctuations in the luminous galaxy distribution, as well as 
deviations from the Hubble flow
of universal expansion known as peculiar velocities.
Optical redshift surveys of galaxies now
examine a 
range of scales out to $\sim 100$ Mpc that 
overlaps with the range
probed by fluctuations in the cosmic microwave background.  

The expected rate of growth of 
density fluctuations 
depends 
on the precise cosmology that is adopted ({\it 2}).  
One can therefore use the comparison between microwave
background anisotropy and fluctuations in the galaxy distribution to
discriminate among rival cosmological models.
Scott, Silk, and White  ({\it 3})
illustrated this comparison.
Several similar analyses ({\it 4-7}) have been presented 
but have used only a portion
of the compilation of observations that we present.

\section{Structure Formation Models}

We examined ten models of structure formation (Table 1), 
which represent the range of 
cosmological parameters currently considered viable ({\it 8}).  
Each model gives transfer functions that predict how 
a primordial power spectrum of infinitesimal
 density perturbations in the early universe develops 
into CMB anisotropies and 
inhomogeneities in the galaxy distribution.
A cosmological model whose predictions agree with 
both types of observations
provides a consistent picture of structure formation on scales ranging from 
galaxy clusters to the present horizon size.  
The cosmological parameter 
$\Omega = \Omega_m + \Omega_{\Lambda}$ gives the ratio
of the energy density of the universe to the critical density necessary to 
stop its expansion.  Critical density is 
$\rho_c = 3 H_0^2 / 8 \pi G$ for a Hubble 
constant of $H_0 = 100h$ km/s/Mpc.  The portion of this critical 
energy density contained in matter is $\Omega_m = \Omega_c + \Omega_\nu
+ \Omega_b$, the sum of the contributions from Cold Dark Matter (CDM), Hot
Dark Matter (HDM) in the form of massive neutrinos, and baryonic matter.  
$\Omega_\Lambda = \Lambda /3 H_0^2$ 
is the fraction of the critical energy density contained
in a smoothly distributed 
vacuum energy referred to as a cosmological constant, $\Lambda$.
  The age of the universe in each model
is determined by the values of $h, \Omega_m$, 
and $\Omega_\Lambda$;
a critical matter density universe has an age of $6.5h^{-1}$Gyr ({\it 9}). 

Each model has a primordial power spectrum of density perturbations 
given by $P_p(k) = A k^n$ where $A$ is the square of a free
 normalization parameter and
$n$ is the scalar spectral index ({\it 10}).  
Scale-invariance ({\it 11}) corresponds to 
$n=1$ for adiabatic (constant entropy) and $n=-3$
 for isocurvature (constant potential) initial density perturbations.
Instead of normalizing to 
the COBE result alone ({\it 12}), 
we found the best-fit normalization of each model (Table 2) using the entire 
data compilation.  
Our 
rationale is that COBE is just one subset of the available 
data, albeit with small error bars, and is in fact
the data most likely to be affected by a possible contribution of gravitational
waves to microwave background anisotropies.  These gravitational
waves from inflation would have a significant
 impact only on large angular scales
and are not traced by the large-scale structure observations.  
Normalizing to all of 
the data 
made our results less 
sensitive to the possible contribution of gravitational waves.  

The first seven models (Table 1) are based on the Standard Cold Dark 
Matter (SCDM) 
model ({\it 13}) and assume that the initial density perturbations
in the universe were adiabatic, 
as is predicted by the 
inflationary universe paradigm.  
The Tilted CDM (TCDM) and Cold + Hot Dark Matter (CHDM) models  
are both motivated by changing 
the shape of the SCDM matter power spectrum to 
eliminate its problem of 
excess power on small scales relative to large scales ({\it 14}). 
The CHDM model has one family of massive neutrinos which contributes
20\% of the critical density ({\it 15}).  
For the cosmological constant ($\Lambda$CDM) 
and open universe (OCDM) models,  
$\Omega_m=0.5, h=0.6$ guarantees roughly 
the right shape of the matter power
spectrum ({\it 5}). 
We have optimized some 
parameters of these models:  $n$ and $\Omega_b$ 
for TCDM, $\Omega_\nu, \Omega_b, n$, and the number of 
massive neutrino families for CHDM, and $\Omega_m, \Omega_b, h$ and 
$n$ for OCDM and $\Lambda$CDM ({\it 16}).    
The $\phi$CDM model ({\it 17}) contains a vacuum energy 
contribution from a late-time
scalar field with $\Omega_\phi = 0.08$.  
This energy behaves like matter today,
but during matter-radiation equality and recombination 
it alters the shape of the 
matter and radiation power spectra from the otherwise
similar SCDM model.  
The Baryonic + Cold Dark Matter (BCDM) model ({\it 18}) contains 
nearly equal amounts of baryonic matter ($\Omega_b = 0.04$) and CDM 
($\Omega_c = 0.08$).     
Its parameters have been tuned to produce a 
peak due to baryonic acoustic oscillations 
in the matter power spectrum at $k=0.05h$Mpc$^{-1}$, 
where a similar peak is seen 
in the 3-dimensional power spectrum of rich Abell clusters ({\it 19}) 
and the
2-dimensional power spectrum of the Las Campanas Redshift Survey 
({\it 20}). 

The Isocurvature Cold Dark Matter (ICDM) 
model ({\it 21, 22}) has a 
non-Gaussian ($\chi^2$) distribution of 
isocurvature 
density perturbations produced by a massive scalar field
frozen during inflation.  This causes 
early structure formation, 
in agreement with observations of galaxies at high 
redshift and the Lyman $\alpha$ forest ({\it 23}).  
The Primordial Black Hole Baryonic Dark Matter
(PBH BDM) model ({\it 24})  
has isocurvature perturbations but no CDM.  The 
primordial black holes form from baryons 
at high density regions in the early 
universe and thereafter behave like CDM.  Only a tenth
of the critical energy density remains outside the black holes to participate
in nucleosynthesis.  These black holes have the appropriate mass ($M \sim
1 M_\odot $) to be
the Massive Compact Halo Objects (MACHOS) which have been detected in 
our Galaxy ({\it 25}). 
Albrecht, Battye and Robinson found 
that critical matter density topological defect models 
fail to agree with structure formation 
observations ({\it 26}).
In the Strings+$\Lambda$ model ({\it 27}) that we examined, 
the nonzero cosmological constant causes a deviation 
from scaling and makes cosmic strings a viable model.

We used the CMBfast code ({\it 28})  
to calculate the predicted radiation and matter power spectra 
for the SCDM, TCDM, CHDM, OCDM, $\Lambda$CDM, and BCDM models.

\section{Constraints on Cosmological Parameters}

The models we consider are all consistent with 
the constraints on the baryon density from Big Bang nucleosynthesis, 
$0.012 < \Omega_b h^2 < 0.026$, 
allowed by recent observations of primordial deuterium abundance ({\it 29}).
  A Hubble constant of $65 \pm 15$ encompasses the range of 
systematic variations between different observational approaches ({\it 30}).
The 
age ``crisis'' has abated with recent recalibration by {\it Hipparcos} of the 
distance to the oldest Galactic globular clusters leading to a new estimate 
of their age of $11.5 \pm 1.3$ Gyr ({\it 31}).  
All of our models have an age of at least 13 Gyrs except OCDM (12 Gyrs).
Other constraints, however, appear to limit the viability of our 
models. 
Observations of high-redshift damped Lyman $\alpha$ systems 
are a concern for the CHDM and TCDM models, 
which have little power at small
scales ({\it 32}).  
Bartelmann {\it et al.} ({\it 33}) used numerical simulations to 
compare the observed abundance of arcs from
strong lensing by galaxy clusters with the predictions of various models
and conclude that only OCDM works, and they found that critical density
models seriously underpredict the number of arcs.
Further support for low-$\Omega_m$ models comes from the 
cluster baryon fraction of  
$\Omega_m/\Omega_b \leq 23 h^{3/2}$ ({\it 34}).
  This favors the ratio of total matter to baryons in 
the low matter density models considered here and is inconsistent with 
SCDM and $\phi$CDM.  
Observations of 
Type Ia supernovae at high redshift 
are progressing rapidly, and preliminary results
argue in favor of a positive cosmological constant and strongly disfavor 
$\Omega_m=1$ ({\it 35}).  
The amount of vacuum energy 
is constrained to be $\Omega_\Lambda \leq
0.7$ by QSO lensing surveys ({\it 36}).  
It is interesting that direct observations of cosmological parameters
favor the low $\Omega_m$ models, but 
we found that the current discriminatory power of observations of 
structure formation outweighs that of direct parameter observations.  

\section{Comparison with Observations}

Since the COBE DMR detection of CMB anisotropy ({\it 1}), there have
been over twenty-five additional measurements of anisotropy on angular scales
ranging from $7^{\circ}$ to $0\fdg3$.
The models predict that the spherical 
harmonic decomposition of the pattern of CMB temperature fluctuations 
on the sky will 
have Gaussian distributed coefficients $a_{\ell m}$ with zero mean 
and variance $C_\ell $.  
Each observation has a window function $W_\ell$ which makes
the total power measured sensitive to a range of angular scales
given by $\theta \simeq 180^\circ / \ell$:

\begin{equation}
\left ( \frac{\Delta T}{T} \right )^2_{rms}
 = \frac{1}{4 \pi} \sum_\ell (2 \ell + 1)C_\ell W_\ell = \frac{1}{2}
(dT/T_{CMB})^2 \sum_\ell \frac {2 \ell + 1}{\ell(\ell+1)} W_\ell , 
\end{equation}

\noindent 
where COBE found $dT = 27.9 \pm 2.5 \mu K$ and
$T_{CMB}=2.73 K$ ({\it 37}).  
This allows the observations of broad-band 
power to be reported as observations of $dT$, and knowing the window
function of an instrument one can turn the predicted $C_\ell$ spectrum
of a model into the corresponding prediction for $dT$ at that
angular scale (Fig. 1).  

We translated these observations of the radiation power spectrum 
 into  estimates of the matter power spectrum
on the same scales ({\it 38}).  
The matter power spectrum is determined by the matter transfer function 
$T(k)$ and primordial power spectrum $P_p(k)$ of each model, with  
$P(k) = T^2(k) P_p(k)$.  
The matter transfer function describes
the processing of initial density perturbations from the Big Bang during
the era of radiation domination; the earlier a spatial scale came within the 
horizon, the more its power was dissipated by radiation 
(and in the CHDM model, by relativistic neutrinos as well).  
If the baryon fraction is large, the same acoustic oscillations of the
photon-baryon fluid that give rise to peaks in the radiation power spectrum
are visible in the matter power spectrum; otherwise, the baryons fall into
the potential wells of the dark matter.    
Once matter domination and recombination arrive, 
$P(k)$ maintains its shape and grows
as $(1+z)^{-2}$.  Thus, determining $P(k)$ today allows us to extract
the power spectrum of primordial density fluctuations that existed when
the universe was over a thousand times smaller.   

Our compilation of observations of fluctuations
in the large-scale distribution of galaxies and galaxy clusters (Fig. 2A)
includes
the determination of $\sigma_8$, the rms density variation 
in spheres of radius $8 h^{-1}$ Mpc, based on 
 the abundance of rich galaxy clusters ({\it 39}).    
Another measurement of $\sigma_8$ 
 is based upon the evolution of the abundance of 
rich clusters from redshift 0.5 until now ({\it 40}).    
  The predicted value of $\sigma_8$
is given by 
an integral over the matter power spectrum using a spherical 
top-hat window function ({\it 41})

\begin{equation}
\sigma_R^2 = \frac{1}{2 \pi^2} \int dk k^2 P(k) \frac{9}{(kR)^6}
(\sin kR - kR \cos kR)^2,  
\end {equation}

\noindent 
which allows observations of $\sigma_8$ to determine the amplitude
of $P(k)$ on scales $k\simeq 0.2 h$Mpc$^{-1}$.    
Another measurement of the amplitude of the  
power spectrum comes from observations of galaxy
 peculiar velocities ({\it 42}). 

Our data compilation includes 
power spectra from four redshift surveys, 
the Las Campanas Redshift Survey (LCRS),  
the combined IRAS 1.2 Jy and QDOT samples, 
the combined SSRS2+CfA2 survey, and a cluster sample selected from 
the APM Galaxy Survey ({\it 43}).
We also use the 
power spectrum resulting from the Lucy inversion of the angular correlation
function of the APM galaxy catalog ({\it 44, 45}).   
The APM galaxy power spectrum is measured in 
real space, whereas the others are given in redshift space. 
Each of these power spectra can be scaled by the square of 
an adjustable bias parameter, which is expected to be near unity 
for the galaxy surveys ({\it 46}).  

Following the methods of Peacock and Dodds ({\it 41}), 
we 
performed model-dependent corrections for 
redshift distortions for each galaxy power spectrum ({\it 47, 48}) 
and divided by the square of a trial value of the bias factor.  
We then corrected for non-linear evolution ({\it 49}) 
to produce estimates of the unbiased linear power spectrum
from these galaxy surveys.  
Comparison with the predicted 
linear $P(k)$ determined the best-fit bias parameter of each survey
for each model (Table 2).  
We compared the corrected large-scale structure data, the CMB 
anisotropy observations and the predicted matter and radiation power spectra 
and calculated the 
$\chi^2$ value for each model (Table 3).  
Only points observed at $k \leq 0.2h$Mpc$^{-1}$ 
were used in selecting best-fit bias factors and normalizations and in 
calculating $\chi^2$ ({\it 50}).  On smaller scales, 
the linearization process yielded 
qualitative information despite systematic uncertainties.

 \section{Discussion}

The current large-scale
structure observations agree well with each other 
 in terms of the shape of the 
uncorrected matter power spectrum (Fig. 2A).  
The APM clusters are biased compared
to galaxies by about a factor of 3 and  
their power spectrum has a narrower peak 
and a possible small-scale feature.
There is no clear evidence,
however, for
scale-dependence in the bias of the various galaxy surveys on linear
scales.  
The observed galaxy power spectra are smooth, showing no
statistically significant oscillations.
A peak in the matter power spectrum appears  
near $k=0.03h$Mpc$^{-1}$, which constrains
$\Omega_m h$ by identifying the epoch of matter-radiation equality 
({\it 44}).  
The large-scale structure observations 
contain too much information 
to be summarized by a single shape parameter; no value of the 
traditional CDM shape parameter ({\it 51}) can simultaneously 
match the location of this peak and its width.  

We find a poor fit for SCDM (Fig. 2B) due to the 
difference in shape between the theory curve and 
the data.  The best-fit normalization is only 0.91 that of COBE, as the 
model would otherwise overpredict the $\sigma_8$ measurements by an 
even greater amount.    
The fit to the CMB is poor, because the Saskatoon (SK) observations 
({\it 52})
would prefer 
more power.  
The fit of the data to the TCDM model (Fig. 2C) 
is better, although the peak of the matter power spectrum 
is still broader than that found in the data.  
Agreement with the CMB is harmed by the high normalization versus COBE 
and the tilt on medium scales.

The best-fit model is CHDM (Fig. 2D).   
The agreement
with the location and shape of the peak of the matter power spectrum
is remarkable, with the exception of the APM cluster power spectrum. 
The agreement
with CMB anisotropy detections is excellent.  
The matter power spectrum of 
CHDM matches the linearized 
APM galaxy power spectrum down to non-linear scales, 
making this model a good explanation of 
structure formation far beyond the scales used for our statistical
analysis ({\it 53}).  

For the OCDM model (Fig. 3A),   
$\Omega_m=0.5$
is favored by the shape of $P(k)$ and 
the SK and CAT ({\it 54}) CMB anisotropy detections
 and generates agreement between the two observations of $\sigma_8$.  
However, the location of the peak of $P(k)$ appears wrong.  
This model is our second-best fit but is 
statistically much worse than CHDM.
The $\Lambda$CDM model (Fig. 3B) is nearly as successful as OCDM.  
It is a slightly better fit to the CMB but is 
worse in comparison to large-scale structure. 
The observations of
$\sigma_8$ are again in agreement, but the shape of the matter power
spectrum does not 
compare well with that of the APM galaxy survey.

The $\phi$CDM model is too broad at the peak 
and misses a number of APM galaxy datapoints (Fig. 3C), although  
its agreement with the other
datasets is rather good.
It remains to be seen whether other 
variations of 
scalar field models can match the observations better.
The BCDM model (Fig. 3D) does not fit the data.  
Choosing parameters to place an acoustic oscillation peak 
near $k=0.05h$Mpc$^{-1}$ has generated the wrong shape for $P(k)$, even
though the APM galaxies and clusters seem to fit the first and second
oscillations, respectively ({\it 55}).    
The main 
peak of $P(k)$ is in the wrong
place; no model with similar oscillations and a baryon content consistent
with Big Bang Nucleosynthesis can fix that problem ({\it 18}). 

For the Isocurvature CDM model (Fig. 4A), 
the fit to the CMB is poor, due 
to the rise of $C_\ell$ on COBE scales, too much power in the first peak 
near $\ell = 100$ and too little power compared to 
SK.  The fit to large-scale structure is mediocre.  
The PBH BDM model has similar problems compared to the CMB,
but the peak location and shape of the matter power spectrum are 
better (Fig. 4B).  
The Strings+$\Lambda$ model (Fig. 4C) underestimates 
the amplitude of the bias-independent measurements and 
therefore requires a large bias
for all types of galaxies, which is difficult to justify.

\section{Conclusions}

The rough agreement of the Cosmic Microwave Background anisotropy and 
Large-Scale Structure observations over
a wide range of models suggests 
that the gravitational instability paradigm of cosmological 
structure 
formation is correct. 
The current set of CMB anisotropy detections may be 
a poor discriminator among adiabatic models, but it strongly prefers
them to non-adiabatic models.  
Several models (SCDM, TCDM, BCDM, and Strings+$\Lambda$) 
have a best-fit 
normalization significantly different from their COBE normalization 
and would 
have been unfairly penalized if normalized to COBE alone.  
The Strings+$\Lambda$ 
model already includes a tensor contribution, but SCDM and BCDM 
would benefit from adding a gravitational wave component that brought
them into better agreement with COBE without changing the amplitude of
their scalar perturbations.  Adding gravitational
 waves is not, however, a panacea
for those models.  In general, the models which are the best fits
to the shape of the matter power spectrum prefer to be close to their
COBE normalization, which argues against there being a significant
tensor contribution to large-angle CMB anisotropies.  

Large-scale structure data 
have more discriminatory power at present 
than do the CMB anisotropy detections.  
The average ratio of best-fit biases (Table 2) is 
$b_{clus}:b_{cfa}:b_{lcrs}:b_{apm}:b_{iras} = 
3.2:1.3:1.2:1.3:1$ ({\it 56}).  
Most models 
allow optical galaxies to be nearly unbiased tracers of the dark 
matter distribution.  The large-scale structure data are smooth enough to 
set a limit on the baryon fraction, $\Omega_b/\Omega_m$; 
when that fraction gets higher
than 0.1 the fit worsens ({\it 57}).

By restricting our analysis to the linear regime 
and correcting for the mildly scale-dependent 
effects of 
redshift distortions and non-linear evolution on those scales, we 
made it possible to test models quantitatively.  
The most likely cosmology 
is Cold + Hot Dark Matter, which is the only model allowed at the 95\% 
confidence level.    
The disagreement between the data and the predictions of the other models is 
sufficient to rule out all of them at above 99\% confidence
unless there
are severe systematic problems in the data ({\it 58}).  
CHDM itself
is not statistically very likely because of the APM cluster
survey $P(k)$, 
which no model fits much better, and which disagrees somewhat 
with the galaxy power spectra.  Dropping the APM 
cluster $P(k)$ would give CHDM a $\chi^2$ 
of 66/62, which is within the 68\% 
confidence interval.
It is worth investigating whether
the APM cluster power spectrum contains a scale-dependent bias 
or if its errors have 
somehow been underestimated.  

We have extracted
the spectrum of primordial density fluctuations from the data and 
found that it agrees well with that of the Cold + Hot Dark Matter
model.
This does
not provide direct evidence for the existence of 
HDM, which requires experimental confirmation of neutrino
mass.  The CHDM model has other observational hurdles to 
overcome, including evidence for early galaxy formation 
on small scales where  
this model has little power, 
although it is impressive that CHDM agrees with the linearized APM 
data out to $k=1h$Mpc$^{-1}$.  
If the rapidly improving 
Type Ia Supernovae observations follow current trends there may 
be enough statistical power in the direct observations of 
cosmological parameters to make OCDM and $\Lambda$CDM 
preferred to CHDM, although in that case 
none of these models 
would be a satisfactory fit to both the supernovae and 
structure formation observations.

\newpage
%\floatsep{0.01 in}
\intextsep 0.02in 
\small
\begin{table}[h] \caption{\small 
Cosmological parameters of our models.  
Parameters marked with a $^*$ were optimized ({\it 59}).}
%\label{tab:models}
\begin{center}
\begin{tabular}{|r|c|c|c|c|c|c|c|c|c|}
\hline
Model &  $\Omega$& $\Omega_{\Lambda}$ & 
$\Omega_{m}$ & $\Omega_c$ & $\Omega_{\nu}$ & 
   $\Omega_{b}$ & h & n & Age (Gyr) \\
\hline
SCDM         & 1.0 & 0   & 1.0   & 0.95   & 0 & 0.05 & 0.5     & 1.0     & 13\\
 
TCDM         & 1.0 & 0   & 1.0   & 0.90  & 0 & $0.10^*$ & 0.5  & $0.8^*$ & 13\\

CHDM	     & 1.0 & 0 & 1.0 & 0.70 & $0.2^*$ & $0.10^*$ & 0.5 & $1.0^*$ & 13\\
 
OCDM       & 0.5 & 0 & $0.5^*$ & 0.45 & 0 & 0.05$^*$ & $0.6^*$ & $1.0^*$ & 12\\

$\Lambda$CDM & 1.0 & 0.5 & $0.5^*$ & 0.45 & 0 & 0.05$^*$ & $0.6^*$ & $1.0^*$ 
& 14\\

$\phi$CDM    & 1.0 & 0   & 0.92  & 0.87   & 0 & 0.05 & 0.5     & 1.0     & 13\\

BCDM         & 1.0 & 0.88 & 0.12 & 0.08   & 0 & 0.04 & 0.8     & 1.6     & 15\\

ICDM         & 1.0 & 0.8 & 0.2   & 0.17   & 0 & 0.03 & 0.7     & -1.8    & 15\\

PBH BDM      & 1.0 & 0.6 & 0.4   & 0    & 0 & 0.10 & 0.7     & -2.0    & 13\\

Strings+$\Lambda$ 
             & 1.0 & 0.7 & 0.3   & 0.25   & 0 & 0.05 & 0.5    & $\sim 1$ & 19\\
\hline
\end{tabular}
\end{center}
\end{table}
\nopagebreak
\begin{table}[h] \caption{\small 
Best-fit normalizations and biases.  The 
normalization of each model is given by $\sigma_8$ or 
the value of $dT$ at $\ell = 10$, which can be compared to 
the COBE normalization of $dT = 27.9 \mu$K.}  
%\label{tab:models}
\begin{center}
\begin{tabular}{|r|c|c|c|c|c|c|c|}
\hline
Model &  $dT_{10}$ ($\mu$K) & $\sigma_8$ & b$_{clus}$ & b$_{cfa}$ & 
   b$_{lcrs}$ & b$_{apm}$ & b$_{iras}$ \\
\hline
SCDM              & 25.4 & 1.08 & 2.12 & 0.83 & 0.72 & 0.89 & 0.57 \\

TCDM              & 31.2 & 0.79 & 2.73 & 1.13 & 1.01 & 1.18 & 0.83 \\

CHDM	          & 27.1 & 0.75 & 2.52 & 1.11 & 1.01 & 1.13 & 0.78 \\

OCDM              & 29.0 & 0.77 & 2.67 & 1.25 & 1.11 & 1.10 & 0.93 \\

$\Lambda$CDM      & 26.8 & 1.00 & 2.14 & 0.91 & 0.82 & 0.87 & 0.68 \\

$\phi$CDM         & 27.6 & 0.74 & 3.12 & 1.35 & 1.20 & 1.31 & 0.98 \\ 

BCDM              & 24.8 & 1.76 & 1.30 & 0.48 & 0.40 & 0.41 & 0.37 \\

ICDM              & 28.2 & 0.83 & 2.95 & 1.25 & 1.12 & 1.02 & 0.97 \\

PBH BDM           & 29.9 & 0.78 & 2.74 & 1.21 & 1.09 & 1.10 & 0.92 \\

Strings+$\Lambda$ & 21.2 & 0.32 & 6.95 & 3.10 & 2.86 & 2.62 & 2.48 \\
\hline
\end{tabular}
\end{center}
\end{table}
\enlargethispage*{1000pt}
\nopagebreak
\begin{table}[h!] \caption{\small
Chi-squared values for our models, computed
from data at $k \leq 0.2h$Mpc$^{-1}$ ({\it 60}).  
P is the probability of getting $\chi^2$ greater than or equal to the 
observed value if a model is correct.}
%\label{tab:models}
\begin{center}
\begin{tabular}{|r|r|r|r|r|r|r|r|r|r|r|}
\hline
Model & $\chi^2_{CMB}$  & $\chi^2_{\sigma_8}$ & $\chi^2_{clus}$ & 
$\chi^2_{cfa}$& $\chi^2_{lcrs}$ & $\chi^2_{apm}$ & $\chi^2_{iras}$ & 
$\chi^2_{total}$ & $\chi^2$/d.o.f. & P~~~~~\\

 d.o.f.  & 34 & 3 & 8 & 2 & 5 & 9 & 9 & 70 &  & \\
\hline
SCDM             & 46 & 36 & 37 & 0.2 & 8 & 121 & 18 & 266 &3.8 & $<10^{-7}$ \\

TCDM             & 51 & 5 & 27 & 0.4 & 6 & 49 & 11  &148  & 2.1 
&$1.8\times 10^{-7}$ \\

CHDM	         & 30 & 4 & 20 & 3 & 9 & 10 & 11 &  86 & 1.2 & 0.09 \\

OCDM             & 36 & 2 & 24 & 2 & 11 & 42 & 12 & 128  & 1.8 
&$2.9\times 10^{-5}$ \\

$\Lambda$CDM    & 30 & 3 & 26 & 2 & 12 & 46 & 13 & 132  & 1.9 
&$1.1 \times 10^{-5}$ \\

$\phi$CDM        & 32 & 4 & 30 & 0.1 & 5 & 71 & 12 & 155  & 2.2 &$<10^{-7}$ \\ 

BCDM             & 32 & 38 & 33 & 1 & 125 & 225 & 56 & 511 & 7.3 &$<10^{-7}$ \\

ICDM             & 61 & 3 & 17 & 2 & 21 & 50 & 16 & 170 & 2.5 & $<10^{-7}$ \\

PBH BDM          & 65 & 4 & 22 & 2 & 9 & 30 & 11 & 142  & 2.0 
&$8.3 \times 10^{-7}$ \\

Strings+$\Lambda$ &64 & 37 & 20 & 0.3 & 8 & 43 & 10 &182 & 2.6 &$<10^{-7}$ \\
\hline
\end{tabular}
\end{center}
\end{table}

\newpage

\normalsize

\begin{figure}
\figurenum{1}
\caption{Compilation 
of CMB anisotropy results with horizontal error bars showing the full width 
at half maximum of each instrument's window function and vertical 
error bars showing the 68\% confidence interval ({\it 61}).   
The detections shown here are COBE, 
FIRS, 
Tenerife, 
South Pole, 
BAM, 
ARGO, 
Python, 
MAX, 
MSAM, 
SK, and
CAT
({\it 62}).
Predictions for the models with their best-fit normalizations 
are plotted as $dT_\ell = (\ell(\ell+1)C_\ell/2 \pi)^{1/2} T_{CMB}$ 
for SCDM (solid black), TCDM (dashed black), CHDM (solid
red), OCDM (dashed blue), $\Lambda$CDM (solid blue), $\phi$CDM (dotted 
black), BCDM (dotted blue), ICDM (dashed magenta), PBH BDM (solid 
magenta), and Strings+$\Lambda$ (dotted magenta).  
The ICDM, PBH BDM, and Strings+$\Lambda$ models disagree with the 
slope implied by 
COBE, SP, and BAM, which prefers the adiabatic models.  
SK favors a high acoustic peak near $\ell = 250$ 
and has small error bars, making it a challenge for most models.  
}
\end{figure}

\begin{figure}
\figurenum{2}
\caption{A.  Compilation of large-scale structure observations 
with $P(k)$ for SCDM (solid curve) shown for reference.  
No corrections for bias, redshift distortions, or non-linear evolution
have been made.  $k$ is the wave number in comoving units of $h$Mpc$^{-1}$. 
The black and blue boxes are measurements of $\sigma_8$ from the present-day
number abundance of rich clusters and its evolution, respectively 
({\it 39, 40}), and the black point with error bars
 is from peculiar velocities 
({\it 42}).  $\Omega_m=1$ is assumed ({\it 63}).   
Uncorrected 
power spectra are shown for the APM galaxy survey (blue triangles), 
Las Campanas (red squares), IRAS (filled pink circles), 
APM clusters (orange circles), 
and SSRS2+CfA2 (green crosses) ({\it 44, 43}).  
B. The SCDM model with its best-fit normalization compared to the 
large-scale structure data with its best-fit biases after 
model-dependent corrections for 
redshift distortions and non-linear evolution ({\it 64}).  
Beyond $k=0.2h$Mpc$^{-1}$, the predicted matter power spectrum
 curve is dotted to indicate uncertainty in the data corrections.  
We plot each CMB anisotropy detection 
as a box, 
where the width of the box represents the range of $k$ to which that
experiment is most sensitive, and the height of the box shows the $68\%$ 
confidence interval ({\it 65}).
C. The TCDM model.  D.  CHDM, our best-fit model.  Note agreement
even on non-linear scales. 
 }
\end{figure}

\begin{figure}
\figurenum{3}
\caption{A. OCDM, with scale-invariance of potential perturbations causing
an increase in the matter power spectrum beyond the curvature scale.  
B. The $\Lambda$CDM model.  C. The $\phi$CDM model.  
D.  The BCDM model.}  
\end{figure}

\begin{figure}
\figurenum{4}
\caption{A.  The ICDM model ({\it 66}).  
B.  The PBH BDM model.  
C.  The Strings+$\Lambda$ model ({\it 67}).
D.  A simulation of 
high-precision 
future observations of CMB anisotropy by the MAP (red boxes) and Planck 
Surveyor (blue boxes) satellites.
Green 
error bars show accuracy of the Sloan Digital Sky Survey and 
magenta are for the 2 Degree Field Survey.   
The simulated data are indistinguishable from the underlying model (CHDM) 
for a wide range of $k$ ({\it 68}).  
}
\end{figure}

%\newpage 

\begin{figure}
\figurenum{5}
\centerline{\psfig{file=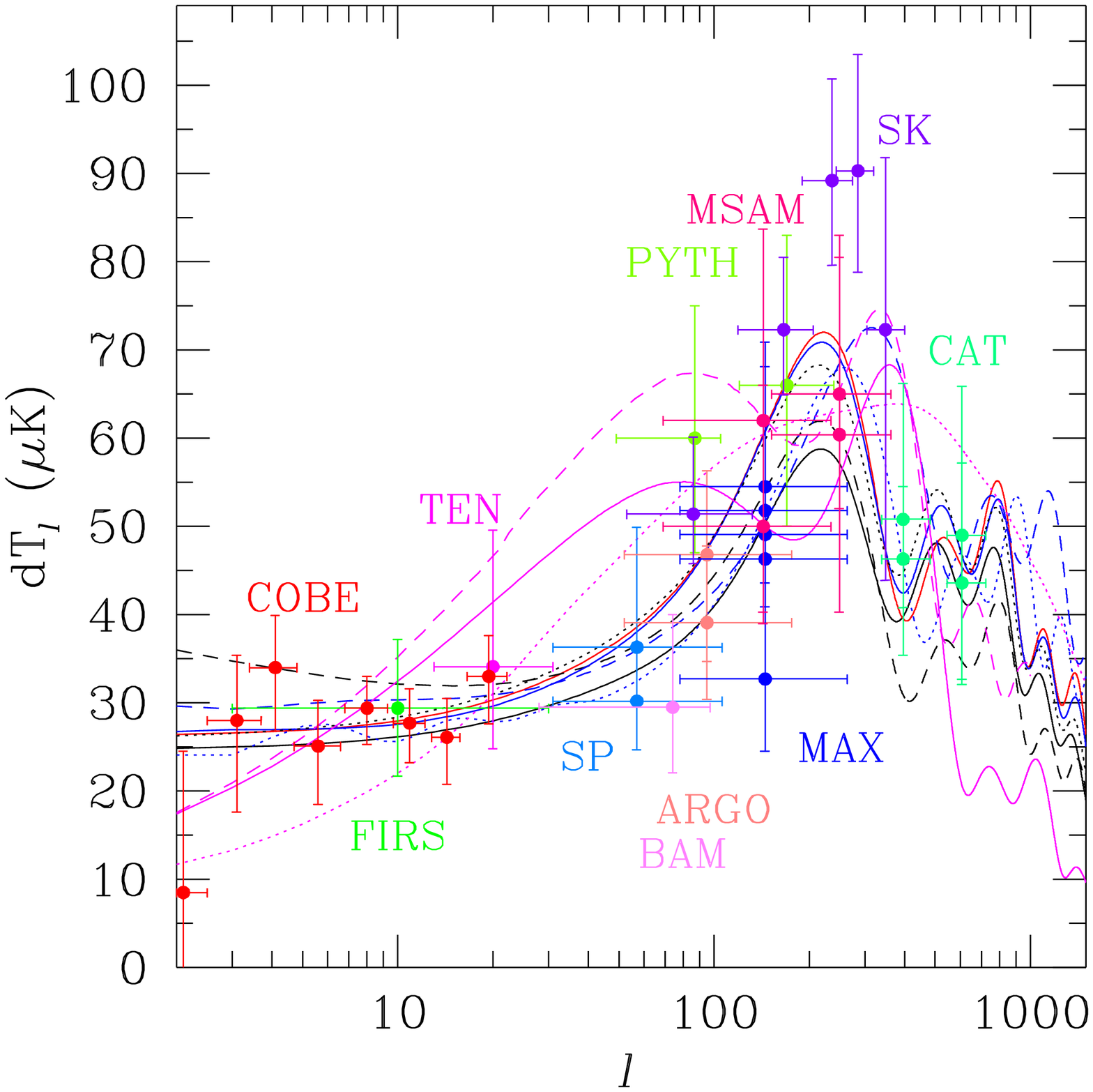}} 
\end{figure}

\begin{figure}
\figurenum{6}
\centerline{\psfig{file=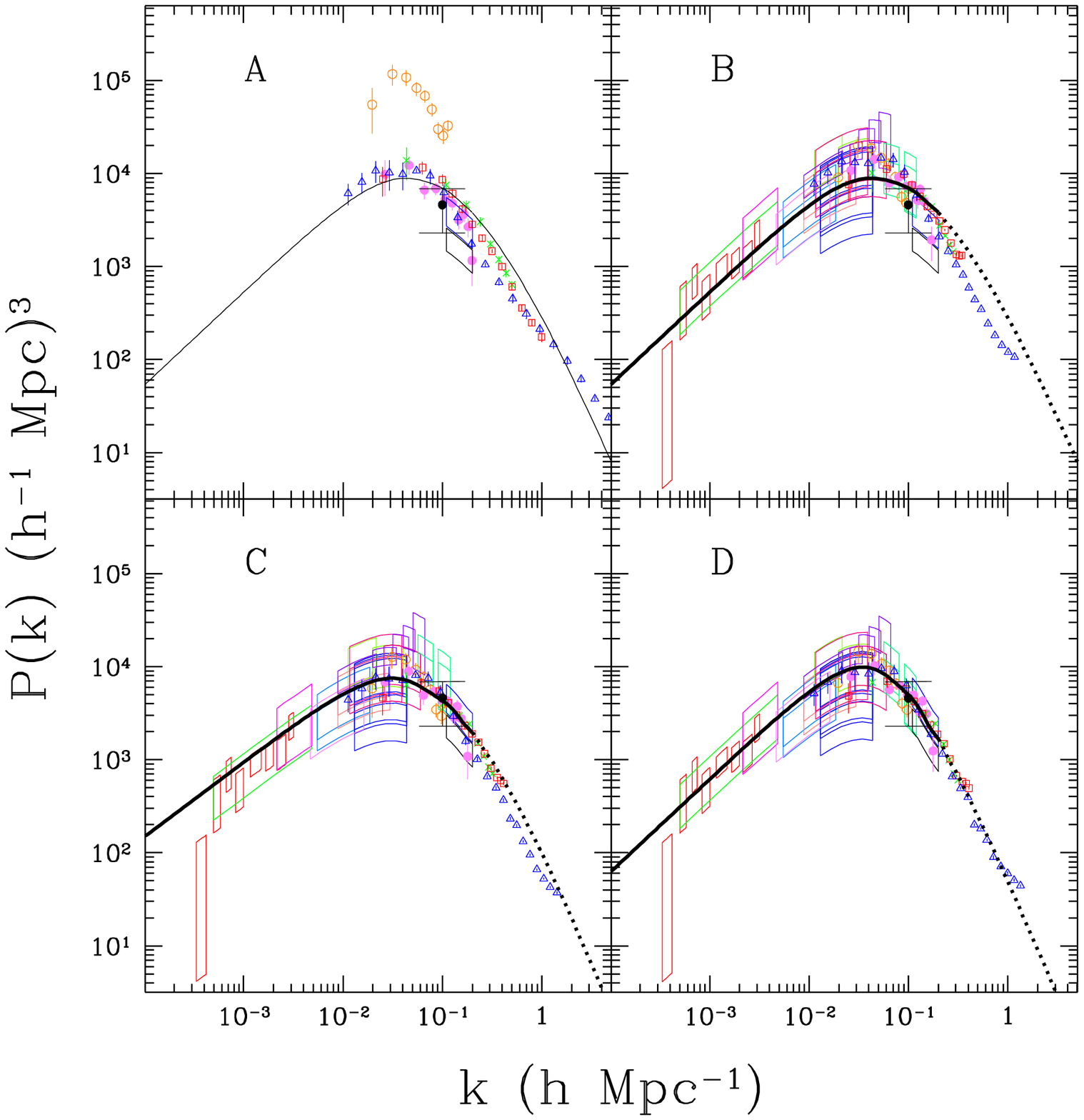}} 
\end{figure}

\begin{figure}
\figurenum{7}
\centerline{\psfig{file=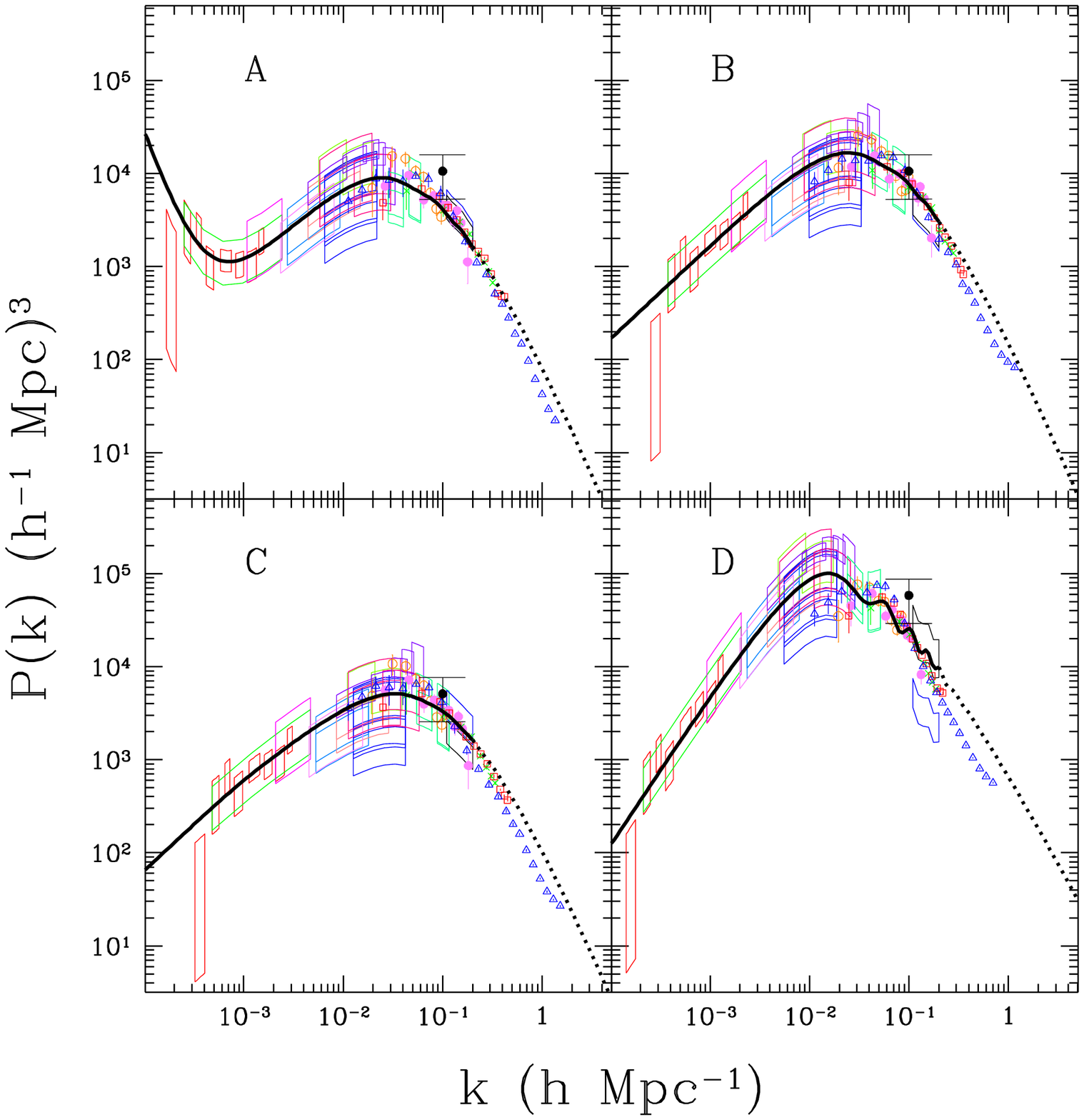}} 
\end{figure}

\begin{figure}
\figurenum{8}
\centerline{\psfig{file=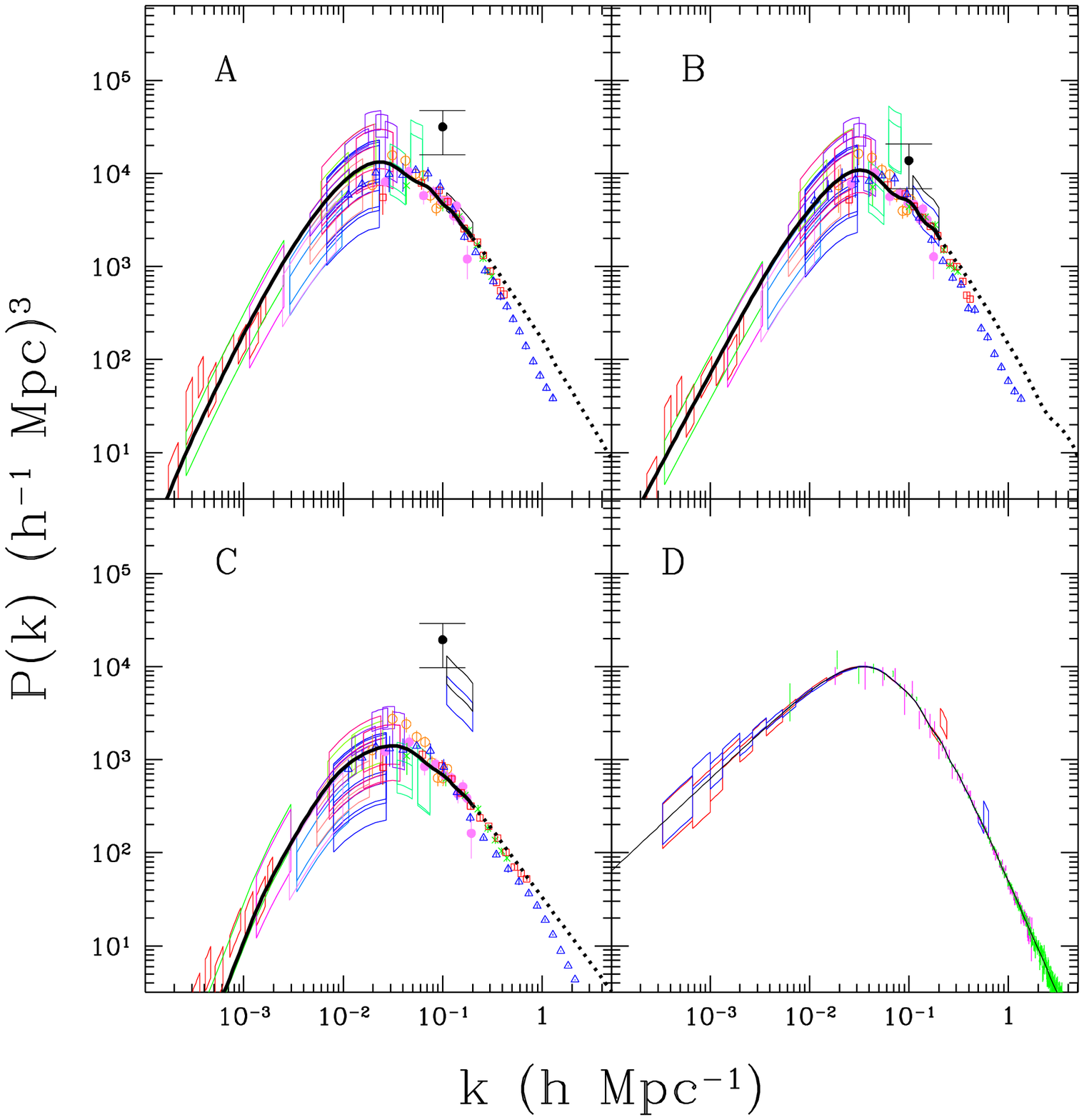}} 
\end{figure}


\begin{references}

\small

%\begin{thebibliography}{}
%\item \ldots

\reference{1}
1.  First reported by the 
COBE DMR team, 
G. F. Smoot {\it et al.}, {\it Astrophys. J.} {\bf 396}, L1 (1992).

\reference{2}
2.  Less fluctuation growth occurs, for
example, in a universe with matter 
density below the critical Einstein-de Sitter value
than in a universe at critical density, because of the weakened role of gravity
at late times.  

\reference{3}
3.  D. Scott, J. Silk, M. White, {\it Science} {\bf 268}, 829 (1995).

\reference{4}
4.   A. N. Taylor and M. Rowan-Robinson, {\it Nature} {\bf 359}, 396 (1992);
M. White and D. Scott, Comments on Astrophysics, Vol. 8, No. 5 
(1996); 
S. Dodelson, E. Gates, M. S. Turner, {\it Science} {\bf 274}, 69 (1996);
S. Hancock, G. Rocha, A. N. Lasenby, C. M. Gutierrez, {\it Mon. Not. 
R. Astron. Soc.} {\bf 289}, 505 (1997);
 J. R. Bond and A. H. Jaffe, in {\it Proceedings of XXXIth Moriond 
Meeting, ``Microwave Background Anisotropies''}, in press (1997); 
C. H. Lineweaver and D. Barbosa, {\it Astrophys. J.} {\bf 496}, 624 (1998); 
 M. Webster, M. P. Hobson, A.N. Lasenby, O. Lahav, and G. Rocha, 
astro-ph/9802109.  Webster {\it et al.} includes a more detailed analysis 
of the galaxy distribution of the IRAS redshift survey
 than is performed here.  When 
we restrict our analysis to the datasets they consider, 
we find that several 
models are in good agreement with the data.

\reference{5}
5.  A. R. Liddle, D. H. Lyth, D. Roberts, P. T. P. Viana, 
{\it Mon. Not. R. Astron. Soc.} {\bf 278}, 644 (1996);
 A. R. Liddle, D. H. Lyth, P. T. P. Viana, M. White, 
{\it Mon. Not. R. Astron. Soc.} {\bf 282}, 281 (1996); 
A. Klypin, J. Primack, J. Holtzman, {\it Astrophys. J.} {\bf 466},
13 (1996).  

\reference{6}
6.  A. R. Liddle, D. H. Lyth, R. K. Schaefer, Q. Shafi, P. T. P. Viana, 
{\it Mon. Not. R. Astron. Soc.} {\bf 281}, 531 (1996).

\reference{7}
7.  M. White, P. T. P. Viana, A. R. Liddle, D. Scott, 
{\it Mon. Not. R. Astron. Soc.} {\bf 283}, 107 (1996).


\reference{8}
8.  Due to theorists' productivity, there are many  
exotic models we could have used, but we included the most popular variations 
of Standard CDM and a sampling of promising alternatives.  


\reference{9}
9.  S. Weinberg, {\it Gravitation and Cosmology} (Wiley, New York, 1972), 
(15.3.11), (15.3.20); S. Carroll, W. Press,  
E. Turner, {\it Ann. Rev. Ast. \& Astrophys.} {\bf 30}, 499 (1992).

\reference{10}
10.  The primordial power spectrum does not have to be an exact power 
law; inflationary models predict slight variation of the power-law 
index with scale.  Allowing this spectral index to vary freely makes it 
more difficult to constrain cosmological models, as shown by 
E. Gawiser and J. Silk, in {\it Proc., Particle Physics and the 
Early Universe Conference}, http://www.mrao.cam.ac.uk/ppeuc/proceedings, 
(1997); E. Gawiser in {\it Proc. of 18th Texas Symp. on Relativistic 
Astrophysics, Chicago, December 1996}, A. Olinto, J. Frieman, D. N. 
Schramm, Eds., in press (Singapore:  World Scientific, 1998).  

\reference{11}
11.  E. R. Harrison, {\it Phys. Rev. D} {\bf 1}, 2726 (1970),
Ya. B. Zel'dovich, {\it Mon. Not. R. Astr. Soc.} {\bf 160}, 1P (1972),
P. J. E. Peebles and J. T. Yu, {\it Astrophys. J.} {\bf 162}, 815 (1970).
For the OCDM
model, we used a primordial power spectrum 
where the scale-invariance of gravitational potential perturbations 
causes a rise in the matter power spectrum at the  
curvature scale.   Critical matter density 
topological defect models have scaling solutions which approximate  
$n=1$.  

\reference{12}
12.  E. F. Bunn and M. White, {\it Astrophys. J.} {\bf 480}, 6 (1997).


\reference{13}
13.  M. Davis, G. Efstathiou, C. S. Frenk, S. D. M. White,  
{\it Astrophys. J.} {\bf 292}, 371 (1985).

\reference{14}
14.  TCDM:  
M. White, D. Scott, J. Silk., M. Davis, {\it Mon. Not. R. Astron. Soc.} 
{\bf 276}, L69 (1995); ({\it 7});  
CHDM:  M. Davis, F. J. Summers, D. Schlegel, {\it Nature} {\bf 359}, 
393 (1992);
A. Klypin, J. Holtzman, J. Primack, E. Regos, {\it Astrophys. J.}
{\bf 416}, 1 (1993);  
D. Y. Pogosyan and A. A. Starobinsky, {\it Mon. Not. R. Astron. Soc.}
{\bf 265}, 507 (1993);
Choi, E. and Ryu, D., {\it Mon. Not. R. Astron. Soc.}, 
in press (1997), astro-ph/9710078.

\reference{15}
15.  The neutrino mass is given by $m_\nu = 94 \Omega_\nu h^2$ 
eV $= 4.7$ eV.


\reference{16}
16.  J. R. Primack, J. Holtzman, A. Klypin, D. O. Caldwell,
{\it Phys. Rev. Lett.} {\bf 74}, 2160 (1995) suggested
 a version of CHDM with two equally massive 
neutrinos to explain the observed cluster abundance, but including 
the higher value of 
$\sigma_8$ implied by the evolution of the cluster abundance 
 makes a single massive neutrino slightly favored. 
We find $\Omega_\nu = 0.2$ preferable to $\Omega_\nu = 0.3$ by a small
margin.  The parameters for CHDM 
and TCDM are within the range found acceptable by ({\it 6}) and ({\it 7}).  

\reference{17}
17.  P. G. Ferreira and M. Joyce, {\it Phys. Rev. Lett.} {\bf 79}, 4740 (1998).
Several other types of scalar fields have been proposed
recently by 
R. R. Caldwell, R. Dave, P. J. Steinhardt, {\it Phys. Rev. Lett.}
{\bf 80}, 1586 (1998); 
P. T. P. Viana and A. R. Liddle, {\it Phys. Rev. D. } {\bf 57}, 674 (1998);
M. S. Turner and M. White, {\it Phys. Rev. D} {\bf 56}, R4439 (1997);
K. Coble, S. Dodelson, J. A. Frieman, {\it Phys. Rev. D} {\bf 55}, 
1851 (1997).  This model is an example that seems to 
agree well with the observations.

\reference{18}
18.  D. J. Eisenstein, W. Hu, J. Silk, A. S.  Szalay, {\it Astrophys. J.} 
{\bf 494}, L1 (1997).    
To reduce the level of CMB anisotropy that results from its scalar 
spectral index of $n=1.6$, BCDM 
has reionization with optical depth to the last scattering surface of 
$\tau = 0.75$.  

\reference{19}
19.  J. Einasto {\it et al.}, {\it Nature} {\bf 385}, 139 (1997).

\reference{20}
20.  S. D. Landy {\it et al.},  {\it Astrophys. J.} {\bf 456}, L1 (1996).


\reference{21}
21.  P. J. E. Peebles, {\it Astrophys. J.} {\bf 483}, L1 (1997).

\reference{22}
22.  P. J. E. Peebles, astro-ph/9805212 (1998).

\reference{23}
23. The numerous atomic hydrogen clouds seen at redshifts $z\sim3$
by their absorption of quasar emission 
are referred to as the Lyman $\alpha$ forest.  They may represent the early
stages of galaxy formation.  

\reference{24}
24.  N. Sugiyama, and J. Silk, in preparation (1998).

\reference{25}
25.  C. Alcock {\it et al.}, {\it Astrophys. J.} {\bf 486}, 697 (1997).

\reference{26}
26.  A. Albrecht, R. A. Battye, J. Robinson,  
{\it Phys. Rev. Lett.} {\bf 79}, 4736 (1997).  
Topological defects are active sources which 
originate in a symmetry-breaking in the early universe and 
seed primordial density perturbations.  
Their
 predictions for matter and radiation fluctuations are similar
to inflationary models, although the matter distribution is 
expected to be non-Gaussian.

\reference{27}
27.  R. A. Battye, J. Robinson, A. Albrecht, {\it Phys. Rev. Lett.}, 
submitted, (1997), astro-ph/9711336.  See also 
P. P. Avelino, R. R. Caldwell, C. J. A. P. Martins, {\it Phys. Rev. D}
{\bf 56}, 4568 (1997).

\reference{28}
28.  U. Seljak and M. Zaldarriaga, {\it Astrophys. J.} 
{\bf 469}, 437 (1996);
M. Zaldarriaga, U. Seljak, E. Bertschinger, {\it Astrophys. J.} {\bf 494},
491 (1998).

\reference{29}
29.  D. Tytler, X.-M. Fan, S. Burles, {\it Nature} {\bf 381}, 207 (1996).  

\reference{30}
30.  D. Branch, {\it Ann. Rev. Ast. \& Astrophys.} {\bf 38} (1998);
W. L. Freedman, in {\it Critical Dialogues in Cosmology}, 
N. Turok, Ed. (World Scientific, Singapore, 1997), pp. 92-129;
A. Sandage and G. A. Tammann, in {\it ibid}, pp. 130-155.

\reference{31}
31.  B. Chaboyer, P. Demarque, P. J. Kernan, L. M. Krauss, 
{\it Astrophys. J.} {\bf 494}, 96 (1998).

\reference{32}
32.  J. P. Gardner, N. Katz, D. H. Weinberg, L. Hernquist, 
{\it Astrophys. J.} {\bf 486}, 42 (1997);
C. P. Ma, E. Bertschinger, L. Hernquist, D. H. Weinberg, N. Katz, 
{\it Astrophys. J.} {\bf 484}, L1 (1997).  The 
improved hydrodynamic simulations of  
M. G. Haehnelt, M. Steinmetz, M. Rauch, (1997) astro-ph/9706201, 
indicate, however, that all 
typical structure formation models have enough small-scale power to 
produce the observations of high-redshift damped Lyman $\alpha$
systems.

\reference{33}
33.  M. Bartelmann, A. Huss, J. M. Colberg, A. Jenkins, F. R. Pearce, 
{\it Ast. \& Astrophys.} {\bf 330}, 1 (1998).

\reference{34}
34.  A. E. Evrard, {\it Mon. Not. R. Astron. Soc.} {\bf 292}, 289 (1997).

\reference{35}
35.  S. Perlmutter {\it et al.}, 
{\it Bulletin of the American Astronomical Society} {\bf 29}, 1351 (1997);  
P. M. Garnavich, et al., {\it Astrophys. J.} {\bf 493}, L53 (1997);   
S. Perlmutter, et al. {\it Nature} {\bf 391}, 51 (1997).

\reference{36}
36.  C. S. Kochanek, {\it Astrophys. J.} {\bf 466}, 638 (1996).  

\reference{37}
37.  C. L. Bennett {\it et al.}, 
{\it Astrophys. J.} {\bf 464}, L1 (1996);
J. C. Mather {\it et al.}, {\it Astrophys. J.} {\bf 420}, 439 (1994).  This 
is based on the standard definition for a ``flat'' power spectrum,
$dT = (\ell (\ell + 1) C_l / 2 \pi )^{1/2} T_{CMB}$.  

\reference{38}
38.  Due to projection effects, 
fluctuations at the last-scattering surface on a given angular scale 
are generated by density perturbations from a range of spatial scales centered
on a wavenumber $k$ (given in units of $h$Mpc$^{-1}$), 
where $\ell \simeq k \eta_0$ and the 
coordinate distance to the last scattering surface is given by 
$\eta_0 = 2 c H_0^{-1} \Omega_m^{-\alpha}$; $\alpha \simeq 0.4$ 
in a flat universe
and $\alpha =1$ in an open universe 
(N. Vittorio and J. Silk, {\it Astrophys. J.} {\bf 385}, L9 (1992)).

\reference{39}
39.  P. T. P. Viana and A. R. Liddle, {\it Mon. Not. R. Astr. Soc.} 
{\bf 281}, 323 (1996).

\reference{40}
40.  N. A. Bahcall, X. Fan, R. Cen, {\it Astrophys. J.} 
{\bf 485}, L53 (1997); X. Fan, N. A. Bahcall, R. Cen, {\it Astrophys. J.} 
{\bf 490}, L123 (1997).  Much attention has recently been paid to using 
the abundance of galaxy clusters at $z=0.2-0.5$ as a probe of $\Omega_m$,
e.g. R. G. Carlberg, S. L. Morris, H. K. C. Yee, E. Ellingson,
{\it Astrophys. J.} {\bf 479}, 82 (1997).  
We accounted for this constraint 
by using both the value of $\sigma_8$ derived from the $z=0$  
cluster abundance and that preferred by the evolution of the 
cluster abundance; models with the preferred value 
of $\Omega_m$ will have these two observations of $\sigma_8$ agree.
  We used the method of 
W. A. Chiu, J. P. Ostriker, M. A. Strauss, {\it Astrophys. J.} {\bf 494},
479 (1998)
to interpret the observed 
cluster abundances and their evolution as measurements of $\sigma_8$ 
for the non-Gaussian 
Strings+$\Lambda$ model.  For ICDM, we used a single
determination of $\sigma_8=0.9 \pm 0.1$ ({\it 22}).

\reference{41}
41.  J. A. Peacock and S. J. Dodds, {\it Mon. Not. R. Astr. Soc.} 
{\bf 267}, 1020 (1994).

\reference{42}
42.  T. Kolatt and A. Dekel, {\it Astrophys. J.} {\bf 479}, 592 (1997).
We used only the
measurement at $k=0.1h$Mpc$^{-1}$  
which is confirmed by the likelihood analysis of 
S. Zaroubi, I. Zehavi, A. Dekel, Y. Hoffman, T Kolatt, 
{\it Astrophys. J.} {\bf 486}, 21 (1997).  A full discussion of peculiar
velocity data is given by M. Gramann, {\it Astrophys. J.} {\bf 493}, 28 (1998).

\reference{43}
43.  Respectively, H. Lin {\it et al.},  
{\it Astrophys. J.} {\bf 471}, 617 (1996);
 H. Tadros and G. Efstathiou, 
{\it Mon. Not. R. Astron. Soc.} {\bf 276}, L45 (1995);
L. N. Da Costa, M. S. Vogeley, M. J. Geller, J. P. Huchra, C. Park,  
{\it Astrophys. J.} {\bf 437}, L1 (1994); 
H. Tadros, G. Efstathiou, G. Dalton, {\it Mon. Not. R. Astron. Soc.}, 
submitted, (1997), astro-ph/9708259.
We used the APM cluster $P(k)$ only for $k\leq 0.12h$Mpc$^{-1}$ 
to avoid possible 
artifacts of the survey window function at higher $k$.    
The APM cluster $P(k)$ was
analyzed for several background cosmologies, and we used the version
most appropriate to each model.
We chose the $101 h^{-1}$Mpc
version of the SSRS2+CfA2 survey to avoid luminosity bias present in 
the deeper sample noted by 
C. Park, M. S. Vogeley, M. J. Geller, J. P. Huchra, 
{\it Astrophys. J.} {\bf 431}, 569 (1994).   

\reference{44}
44.  E. Gazta\~{n}aga and C. M. Baugh, {\it Mon. Not. R. Astron. Soc.}
{\bf 294}, 229 (1998).
We dropped 
the first four reported APM points because they appear to be artifacts
of the data analysis.  

\reference{45}
45.  C. M. Baugh and G. Efstathiou, {\it Mon. Not. R. Astr. Soc.} 
{\bf 265}, 145 (1993).

\reference{46}
46.  
Galaxies may be more or less common than $1\sigma$ peaks
of the dark matter distribution, so they are biased tracers of the mass.  
Each morphological type 
is expected to have a slightly different 
scale-independent bias. Clusters have a large bias
because they trace high density peaks of the primordial density 
distribution and such peaks are themselves highly clustered (see 
N. Kaiser, {\it Astrophys. J.} {\bf 284}, L9 (1984)).   
Our assumption that bias is scale-independent
is supported by G. Kauffmann, A.
Nusser, M. Steinmetz, {\it Mon. Not. R. Astron. Soc.} {\bf 286}, 795 (1997);
R. G. Mann, J. A. Peacock, A. F. Heavens, {\it Mon. Not. R. Astron. Soc.} 
{\bf 293}, 209 (1998);
and R. J. Scherrer and D. H. Weinberg, (1997), astro-ph/9712192.   
Peculiar velocities arise due to the gravity of the underlying 
dark matter, so they produce a bias-independent measurement of the 
matter power spectrum.
Both observations of $\sigma_8$ 
are also bias-independent.  

\reference{47}
47.  The power spectrum observed
in redshift space is related to that in real space by 
$ P_z(k) = (1 + \beta \mu^2)^2 D(k \mu \sigma_p) P_{real}(k)$, 
where the first term gives the Kaiser distortion 
(N. Kaiser, {\it Mon. Not. R. Astr. Soc.} {\bf 227}, 1 (1987)) 
 from coherent infall
of galaxies with bias $b$ as a function of $\beta = \Omega_m^{0.6} / b$ 
and the second term is the 
damping of such distortions by the rms pairwise galaxy velocity dispersion 
($\sigma_p$) measured in units of $H_0$.  This velocity dispersion leads to
the so-called fingers-of-God effect in redshift surveys.
For an exponential velocity distribution,
$D(k \mu \sigma_p) = (1 + (k \mu \sigma_p)^2/2)^{-1}$.  
We averaged over $\mu$, the cosine of the 
angle between the line of sight and a given wave vector $\mathbf{ k}$, 
to produce
an estimate of the real-space power spectrum  
$P_{real}(k) = P_z(k) / f(k,b)$.  Defining $K = k \sigma_p / \sqrt{2}$, 
this gave (W. E. Ballinger, Thesis, University of Edinburgh, 1997):

\begin{equation}
f(k,b) = \frac{1}{K}\left [ \tan^{-1}(K) \left (1 - \frac{2 \beta}{K^2} + 
\frac{\beta^2}{K^4} \right ) + \frac{2 \beta}{K} + \frac{\beta^2}{3 K} - 
\frac{\beta^2}{K^3} \right ].
\end{equation}

For the pairwise velocity dispersion, we used the observation 
from S. D. Landy, A. S. Szalay, T. J. Broadhurst, {\it Astrophys. J.}
{\bf 494}, L133 (1998)
of $\sigma_p = 3.63 h^{-1}$Mpc.  Their determination that 
the velocity distribution is exponential motivated using that form for  
the damping term.  
We tried using the higher value of $\sigma_p = 5.70 h^{-1}$Mpc 
(Y. P. Jing, H. J. Mo, G. Borner, {\it Astrophys. J.} {\bf 494}, 
1 (1998)),  
and it made only a small difference on quasi-linear
scales; at $k=0.2h$Mpc$^{-1}$ the scale-dependence of the redshift
distortions is a 15\% effect for the value of $3.63 h^{-1}$Mpc 
and twice that
for the higher one. This systematic uncertainty in the pairwise 
velocity dispersion makes the corrected real-space power spectrum somewhat 
unreliable on smaller scales.
For the cluster power spectrum, 
we used only the 
scale-independent Kaiser distortion term to correct for redshift distortions
as the clusters are engaged in coherent infall onto superclusters.     
The APM galaxy power spectrum was corrected
for bias only, as it does not have redshift distortions.  
 



\reference{48}
48.  C. Smith, A. Klypin, M. Gross, J. Primack, J. Holtzman,  
{\it Mon. Not. R. Astron. Soc.}, submitted (1997), astro-ph/9702099  
give a full discussion of the effects of varying 
$\sigma_p$ and propagating this systematic uncertainty through the 
linearization procedure; their results confirm that the systematic 
uncertainty is small up to $k=0.2h$Mpc$^{-1}$.  


\reference{49}
49.  Because collapsing structure leads to 
a change of physical scale, the 
observed $k_{nl}$ can be corrected to their
linear values, given by
$k_l = (1 + \Delta^2_{nl})^{-1/3}k_{nl}$, where 
$\Delta^2=k^3P(k)/2\pi^2$.  
The non-linear evolution is given by 
$\Delta^2_{nl} = f(\Delta^2_l)$.  A semi-analytic fit for this 
function with 10\% accuracy compared to numerical simulations is given 
by J. A. Peacock and S. J. Dodds, {\it Mon. Not. R. Astron. Soc.} 
{\bf 280}, L19 (1996).  The accuracy of this formula is confirmed in  
({\it 48}).  This correction is model-dependent, as it assumes a local 
slope for the original 
linear power spectrum based on the model being tested.  
By inverting the formula numerically,
we linearized the unbiased real-space $P(k)$.
Non-linear evolution is significant only at 
$k\geq 0.2h$Mpc$^{-1}$.  At smaller $k$,  
the linearization preserves the shape of the observed 
non-linear $P(k)$ while sliding the data points to smaller $k$;
its main effect is to shrink the error bars slightly.


\reference{50}
50.  We have rebinned some of the galaxy survey data
to make the points independent.  
Our conclusions are unaffected by varying the non-linear 
cutoff between
$k=0.15h$Mpc$^{-1}$ and $k=0.25h$Mpc$^{-1}$.  

\reference{51}
51.  Defined by G. Efstathiou, J. R. Bond, S. D. M. White, {\it 
Mon. Not. R. Astron. Soc.} {\bf 258}, 1P (1992).  

\reference{52}
52.  C. B. Netterfield {\it et al.}, {\it Astrophys. J.} {\bf 474}, 47 (1997).
We used the recalibration of SK from E. Leitch, Thesis, 
Caltech, (1997).

\reference{53}
53.  J. A. Peacock, {\it Mon. Not. R. Astr. Soc.} {\bf 284}, 885 (1997) 
and ({\it 48}) also found good agreement 
between the linearized observations and linear theory under the CHDM 
model.

\reference{54}
54.  P. F. S. Scott {\it et al.}, {\it  Astrophys. J.} 
{\bf 461}, L1 (1996); J. C. Baker, in 
{\it Proc. Particle Physics and the Early Universe Conference}, (1997),
http://www.mrao.cam.ac.uk/ppeuc/proceedings.  

\reference{55}
55.  We averaged the 
predictions of the matter power spectrum over the window function of the 
observations to take into account the possible smoothing of 
these oscillations during observation.  
To make the linearization procedure work smoothly, we fixed
the local slope of the linear power spectrum.


\reference{56}
56.  This is roughly 
consistent with the bias ratios found by ({\it 41}).  

\reference{57}
57. See D. M. Goldberg and M. A. Strauss, {\it Astrophys. J.} {\bf 495},
29 (1998) for the future prospects of this constraint.

\reference{58}
58.  
If the APM Galaxy $P(k)$ were ignored, 
OCDM, $\Lambda$CDM, and $\phi$CDM
would become much better fits but would 
still be ruled out at 95\% confidence.  

\reference{59}
59.  The $\phi$CDM model has a scalar field 
energy density of $\Omega_\phi = 0.08$. 
The PBH BDM model has 30\% of critical density in primordial 
black holes which act like CDM but actually contain baryonic matter.  


\reference{60}
60.  
The $\chi^2_{\sigma_8}$ category includes the contribution from peculiar
velocity measurements.  The degree of freedom used by normalizing is
counted under $\chi^2_{CMB}$, and each galaxy survey loses one 
degree of freedom in choosing a best-fit bias.  The ICDM model has one less
degree of freedom in the $\chi^2_{\sigma_8}$ column and a total of 69.


\reference{61}
61.  The error bars include 
uncertainties due to instrument noise,  
calibration uncertainty, sample
variance from observing only part of the sky, and cosmic variance
from observing at only one location within the universe.  
The calibration errors were 
added in quadrature.  Although calibration errors 
are correlated for multiple observations
by the same instrument, they have been treated as independent, which is
a good approximation after 
the recalibration of SK by Leitch ({\it 52}).  


\reference{62}
62.  CMB anisotropy 
observations are compiled in G. F. Smoot and D. Scott, 
in {\it Review of Particle Properties}, (1998), 
astro-ph/9711069 and at http://www.sns.ias.edu/\~~max/cmb/experiments.html. 
Shown here are COBE (M. Tegmark and A. Hamilton, (1997) astro-ph/9702019),
FIRS (K. Ganga, L. Page, E. Cheng., S. Meyers, {\it Astrophys. J.} 
{\bf 432}, L15 (1993)),
Tenerife (C. M. Gutierrez, {\it et al.}, 
{\it Astrophys. J.} {\bf 480}, L83 (1997)), 
South Pole (J. O. Gundersen {\it et al.}, {\it Astrophys. J.} 
{\bf 443}, L57, (1994)), 
BAM (G. S. Tucker {\it et al.}, {\it Astrophys. J.} {\bf 475}, L74 (1997)),  
ARGO (S. Masi {\it et al.}, {\it  Astrophys. J.} {\bf 463}, L47 (1996)), 
Python (S. R. Platt {\it et al.}, {\it  Astrophys. J.} {\bf 475}, L1 (1997)), 
MAX (M. Lim {\it et al.}, 
{\it  Astrophys. J.} {\bf 469}, L69 (1996); S. T. Tanaka 
{\it et al.}, {\it  Astrophys. J.} {\bf 468}, L81 (1996)),
MSAM (E. S. Cheng {\it et al.}, {\it  Astrophys. J.} {\bf 488}, L59  
(1997)),
SK ({\it 52}), and  
CAT ({\it 54}).
%WD (G. S. Tucker, G. S. Griffin, H. T. Nguyen, J. B. Peterson,  
%{\it  Astrophys. J.}
%{\bf 419}, L45 (1993)),
%OVRO (A. C. S. Readhead {\it et al.}, 
%{\it  Astrophys. J.} {\bf 346}, 566 (1989)),
%SUZIE (S. E. Church {\it et al.}, {\it  Astrophys. J.} {\bf 484}, 523 (1997)),
%ATCA (R. Subrahmayan, R. D. Ekers, M. Sinclair, J. Silk, {\it
%Mon. Not. R. Astr. Soc.}, {\bf 263}, 416 (1993)),
%VLA (B. Partridge {\it et al.}, {\it  Astrophys. J.} {\bf 483}, 38 (1997).

\reference{63}
63.  The 
width of the box represents the range of spatial scales to which $\sigma_8$ is
sensitive and the height shows the $68\%$ confidence interval.  
The half-max 
window for $\sigma_8$ is from $k=0.05$ to $k=0.3$ but it 
has been narrowed for clarity.  The width of the window function 
of the peculiar
velocity observation is shown by the ends of its error bars, 
which include cosmic variance.  
This observation scales as $\Omega_{m}^{-1.2}$ (the square of the growing
mode).  
The determination of $P(k)$ from the 
value of $\sigma_8$ implied by the $z=0$ 
cluster abundance  
scales roughly as $\Omega_m^{-1}$ due to the relationship 
between the observed mass and the pre-collapse radius of rich clusters.  
The observation of $\sigma_8$ from the evolution of the cluster
abundance is nearly independent of $\Omega_m$.

\reference{64}
64.  The high-$k$  
end of the LCRS data shows that the combination of deconvolving the 
fingers-of-God and linearizing the data has kept the shape
the same but moved the points along that curve and 
reduced the error bars.  The linearization
of the APM dataset has removed the inflection
 at $k=0.2h$Mpc$^{-1}$.

\reference{65}
65.  Each model has a particular value of $\eta_0$; varying $\eta_0$  
moves the entire set of boxes horizontally.
Comparison of the CMB anisotropy predictions of each model 
with observations gives the vertical placement of each box, 
showing
the inferred amplitude of matter density fluctuations at that scale.
  The boxes for CMB anisotropy detections and $\sigma_8$ 
follow the local shape of each model's $P(k)$ to 
indicate that they 
are a model-dependent averaging of the power over a range of $k$.  


\reference{66}
66.  The strong 
rise of the matter power spectrum is caused by the sharp tilt of the model
away from scale-invariance.  
The linearization procedure used was calibrated 
for Gaussian models, but non-linear evolution is expected to be 
similar under the $\chi^2$ distribution; see A. Stirling, Thesis,
University of Edinburgh, 1998.  

\reference{67}
67.  It is possible that the linearization
procedure needs to be adjusted to account for non-Gaussianity
in the matter distribution, but 
the reduced power of this
model weakens the effects of non-linear evolution.       


\reference{68}
68.  MAP and Planck parameters taken from J. R. Bond, G. 
Efstathiou, M. Tegmark, {\it Mon. Not. R. Astr. Soc.} 
{\bf 291}, L33 (1997).  SDSS is from 
M. Vogeley, personal communication (1997), and 2DF is from 
S. Hatton, personal communication (1997).  
No attempt has been made in this figure to account
for redshift distortions 
or non-linear evolution.  
The overlap in scale between CMB anisotropy detections and 
large-scale structure observations will increase tremendously in the
next several years, and the errors in these measurements will decrease
significantly.     
W. Hu, D. J. Eisenstein, M. Tegmark, submitted to {\it Phys. Rev. Lett.},
(1998), astro-ph/9712057 
examine how well $\Omega_\nu$ can be determined by SDSS observations,
and Y. Wang, D. N. Spergel, M. A. Strauss, submitted to 
{\it Astrophys. J.}, (1998), astro-ph/9802231 
 discuss the ability of combined MAP and SDSS observations to 
constrain cosmological parameters.  

\reference{69}
69.  
We gratefully acknowledge the contributions of U. Seljak and M. Zaldarriaga 
(CMBFAST), P. Ferreira ($\phi$CDM), J. Peebles (ICDM), N. Sugiyama (PBH BDM),
J. Robinson (Strings + $\Lambda$), H. Tadros (APM cluster analysis 
for various cosmologies), and the efforts of hundreds of observers
who gathered the data.  This work benefitted from conversations with
H. Tadros, M. White, M. Vogeley, P. Viana, A. Taylor, 
A. Stirling, J. Robinson, J. Peebles,
J. Peacock, A. Liddle, A. Jaffe, W. Hu, A. Heavens, G. Evrard, C. Baugh, 
and W. Ballinger.  We thank L. Moustakas for comments on a draft of this 
paper.  
E. Gawiser acknowledges the support of an NSF Graduate Fellowship.  We thank
the Institut d'Astrophysique de Paris for hospitality during the completion
of this research.    

\vspace{1 in}
{\large Our data compilation and full-size figures are available at
{\bf http://cfpa.berkeley.edu/cmbserve/fluctuations/figures.html}}.  

\end{references}
\end{document}